\documentclass[a4paper,11pt]{article}
\usepackage{pos}
\usepackage{bm}
\usepackage{comment}
\usepackage{bbm}
\usepackage{multirow}
\usepackage{subcaption}

\usepackage{pos}
\usepackage{hyperref}
\usepackage{lipsum} 

\usepackage{setspace}

\usepackage{tikz}
\usetikzlibrary{shapes, positioning}

\manuallySeparateAuthors

\title{VAE-based latent-space classification of RNO-G data}
 \ShortTitle{VAE classification of RNO-G data}

\author*[a]{Thorsten Gl\"usenkamp}
\author{ {for the RNO-G Collaboration}\\{\normalsize \normalfont(a complete list of authors can be found at the end of the proceedings)}\\}
\affiliation[a]{Uppsala University, Lägerhyddsvägen 1, 751 20 Uppsala, Sweden}

\emailAdd{thorsten.gluesenkamp@physics.uu.se}

\abstract{The Radio Neutrino Observatory in Greenland (RNO-G) is a radio-based ultra-high energy neutrino detector located at Summit Station, Greenland. It is still being constructed, with 7 stations currently operational. Neutrino detection works by measuring Askaryan radiation produced by neutrino-nucleon interactions. A neutrino candidate must be found amidst other backgrounds which are recorded at much higher rates---including cosmic-rays and anthropogenic noise---the origins of which are sometimes unknown. Here we describe a method to classify different noise classes using the latent space of a variational autoencoder. The latent space forms a compact representation that makes classification tractable. We analyze data from a noisy and a silent station. The method automatically detects and allows us to qualitatively separate multiple event classes, including physical wind-induced signals, for both the noisy and the quiet station.}

\ConferenceLogo{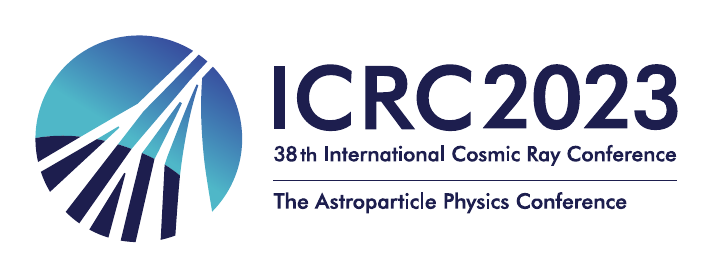}

\FullConference{%
38th International Cosmic Ray Conference (ICRC2023)\\
  26 July - 3 August, 2023\\
  Nagoya, Japan}

\begin{document}
\maketitle

\section{Introduction}

In the last decade, the field of high-energy neutrino astronomy has been established. At energies between 100 GeV and a few PeV, astrophysical neutrinos have been detected as extragalactic diffuse flux\cite{hese_refined}, from the galaxy as galactic diffuse emission\cite{galactic_plane_results} and from a few AGN as point source emission\cite{pointsources}.  However, because the neutrino flux is steeply falling towards higher energies, current detectors like IceCube are not able to detect significant amounts of neutrinos above $10$ PeV over its lifetime, let alone $100$ PeV. The only viable way is to build a detector with a much bigger effective volume. One such effort is RNO-G (Radio Neutrino Observatory Greenland)\cite{rnog_whitepaper}, which is currently being constructed in Greenland. It measures high energy neutrinos via Askaryan radiation originating from the particle cascade following their interactions in the ice. The threshold for detection is about $100$~PeV and extends into the EeV range, which allows it to complement current neutrino facilities to the highest energies, in particular for cosmogenic neutrino fluxes\cite{cosmogenic_neutrinos}. 

In order for RNO-G to make discoveries, it is vital to detect and understand different types of background events. In this contribution, we study the unsupervised detection of event classes in the shallow antennas of a RNO-G station in a 2-step process. First we, compress the data into a latent representation using a variational autoencoder (VAE)\cite{vae_paper}. Second, we cluster the data in the latent space and search for distinct classes. For the study we use the shallow component of two stations: station 21, which has more noise from electronics and is closer to human activity, and station 23, which has less electronically induced noise and is more remote.

\section{Station and data}

RNO-G currently consists of seven stations, spread out within a few kilometers close to Summit Station, Greenland. A single station consists of 24 antennas, of which nine are located just below the surface. In this contribution, we will only focus on the shallow antennas. These are log-periodic dipole arrays (LPDAs), which are spread out in groups of three around the central data acquisition box. Three of the nine antennas are upwards facing, and six are downward facing. A threshold trigger algorithm is run separately on the downward facing and upward facing antennas, whose threshold is tuned to be 1 Hz. Once a trigger happens, an event consists of a 2048 samples long antenna trace for each antenna with a time resolution of roughly 0.3 ns.  If there is more noise in the environment, the threshold is increased to keep the trigger rate constant. Additionally, there are fixed readouts of the whole detector in 10 second intervals. We focus on two stations. The first is station 21, which we call "noisy" station, and is located about one kilometer from the summit camp. The second is station 23, which we call "silent" station, as it is located about 4 km away from summit camp. The noisy station triggers significantly more events from human interference, and is also sensitive to background from battery charging and a long range wide area network (LoRaWAN) noise. 

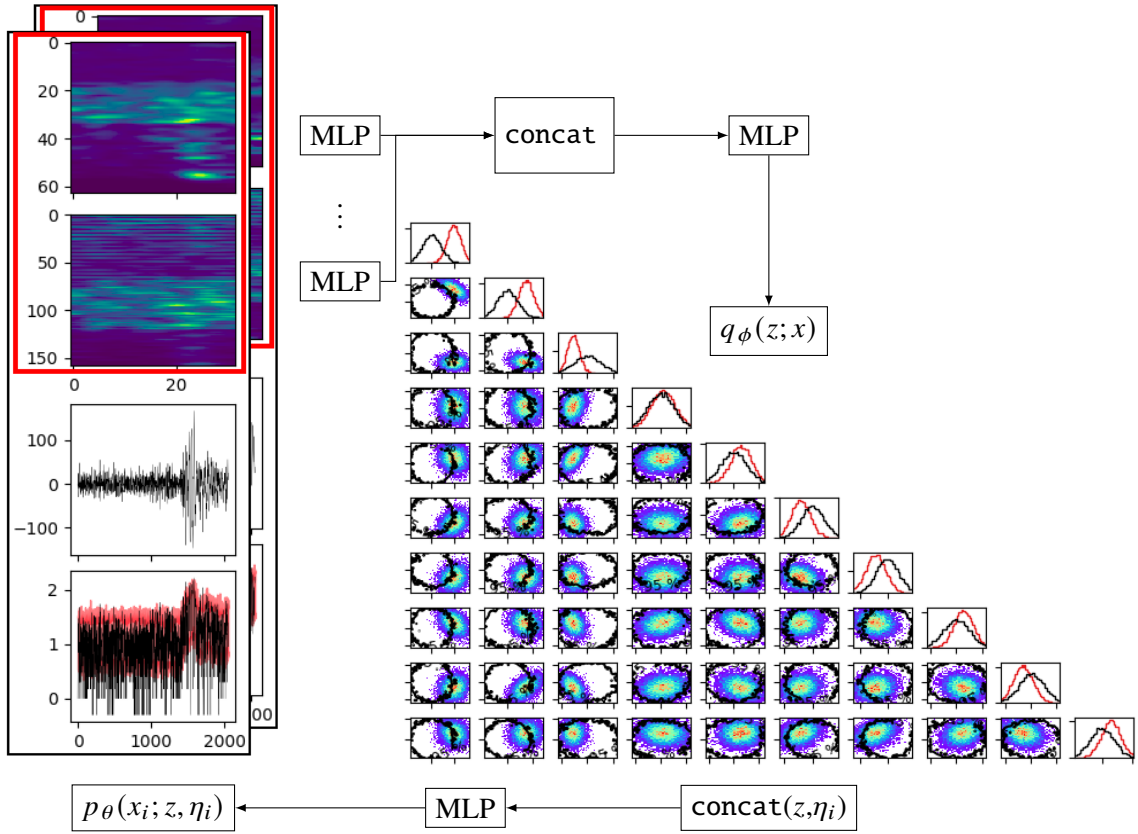
\begin{figure}[h!]
\centering
\begin{tikzpicture}[node distance=1.5cm, >=latex]

  \node (input) at (0, 0) {\includegraphics[width=4cm]{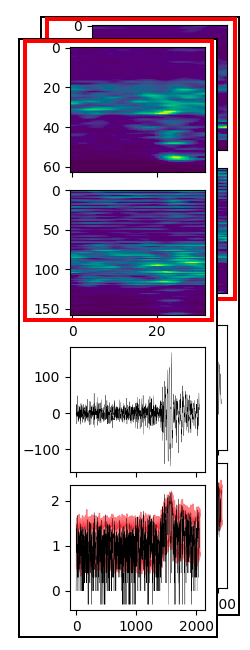}};
  

  \node [ right=0.4cm of input] (input5) {};
  
  
  \node [ above=0.3cm of input5] (input4) {};
  \node [draw, rectangle, above=0.3cm of input4] (input3) {MLP};
  \node [ above=0.3cm of input3] (input2) {$\vdots$};
  \node [draw, rectangle, above=0.3cm of input2] (input1) {MLP};

  \node [draw, rectangle, right=1.5cm of input1, minimum height=1cm, minimum width=1cm,, align=center] (flatten) {\texttt{concat} };

  \node [draw, rectangle, right=1.5cm of flatten] (second_mlp) {MLP};

  \foreach \i in {1,3}
    \draw [->] (input\i.east) -- ([xshift=0.2cm]input\i.east) |- ([yshift=0.0cm]flatten.west);

  \draw [->] (flatten.east) -- (second_mlp.west);

  \node [below=0.7cm of second_mlp](output) {\includegraphics[width=9.7cm]{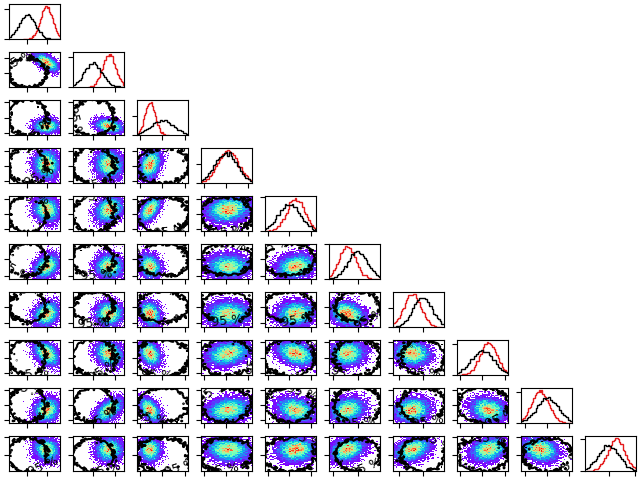}};

  \node [draw, rectangle, below=0.1cm of output] (concat) {\texttt{concat}($z$,$\eta_i$)};

   \node [draw, rectangle, below=2cm of second_mlp] (q_string) 
  {$q_\phi(z;x)$};
   \node [draw, rectangle, left=2.3cm of concat] (third_mlp) {MLP};

   \node [draw, rectangle, left=2.5cm of third_mlp] (llh) {$p_{\theta}(x_i;z,\eta_i)$};
   
  \draw [->] (second_mlp.south) -- (q_string.north);
  \draw [->] (concat.west) -- (third_mlp.east);
  \draw [->] (third_mlp.west) -- (llh.east);
\end{tikzpicture}
\caption{Overview of the VAE. The left part indicates a summary of antenna data: the upper two plots (red boxes) show the scattering coefficients of the voltage trace (see text), the lower two show the voltage trace of the antenna in linear scale and as the logarithm of the absolute value. The processing of the data moves clockwise following the arrows: the scattering coefficients for each antenna $i$ are fed to a multi-layer perceptron (MLP) that is shared among all antennas. The output of all MLPs is concatenated and fed into a second MLP that predicts the parameters of the 10-d latent space posterior distribution $q_\phi(z;x)$. An example posterior distribution is shown by visualizing its probability mass projection in 2-d (histogram) and 1-d (red), i.e. plotting its marginal distributions. The prior marginal distributions are also shown (black) and in 2-d are indicated by their $95 \%$ contour. The data distribution $p_\theta(x_i; z,\eta_i)$ for a given antenna $i$ is predicted by another MLP that takes as input a latent coordinate $z$ concatenated with antenna orientation parameters $\eta_i$. The $95~\%$ envelope of samples from $p_\theta(x_i; z,\eta_i)$ is indicated in red on top of the logarithm of the absolute value of the data trace, which is the target space used for prediction.}
\label{fig:vae_pipeline}
\end{figure}

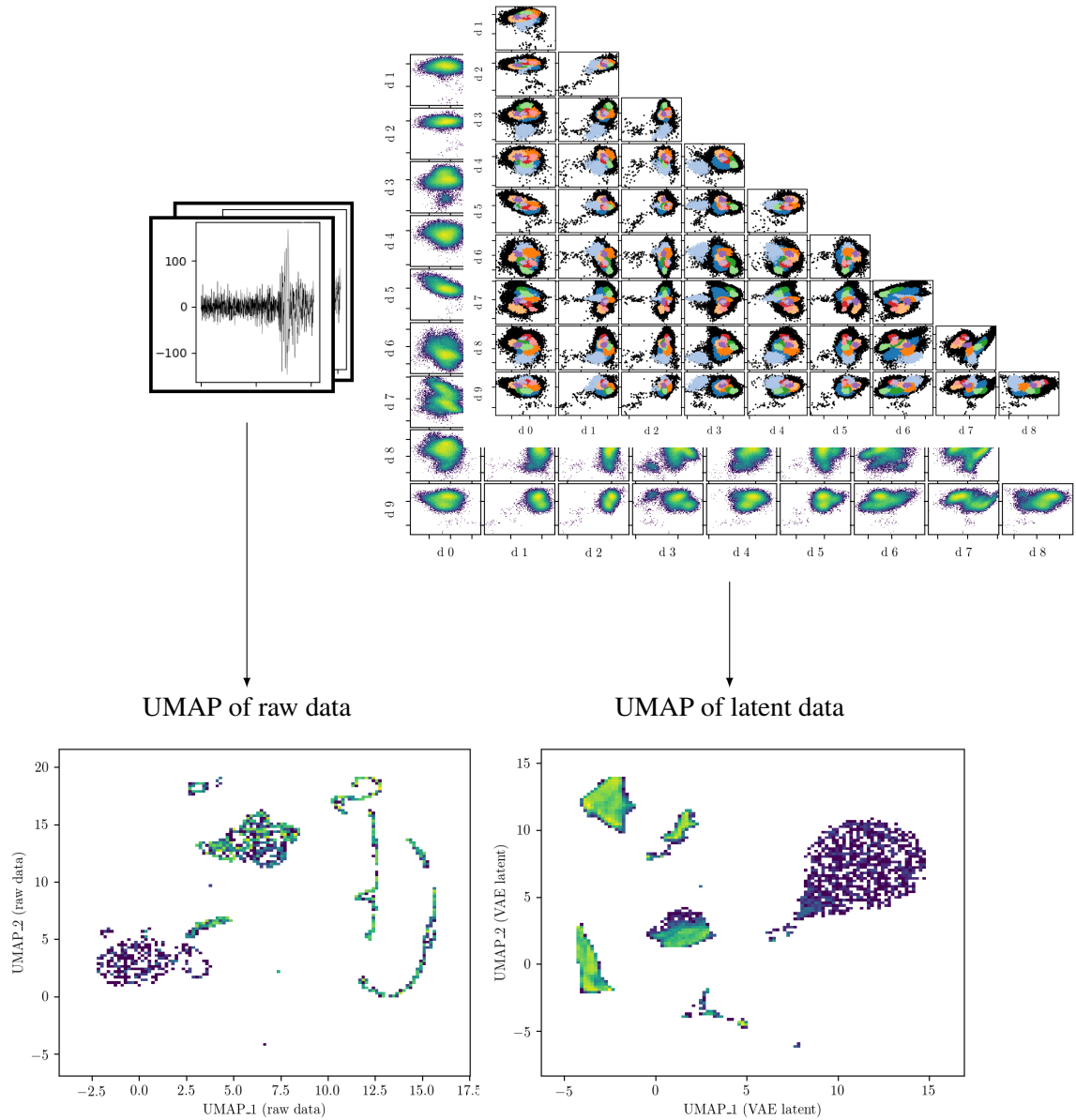
\begin{figure}
\centering
\begin{tikzpicture}[node distance=1.5cm, >=latex]

   \node (raw_antenna_traces) at (0, 0) {\includegraphics[width=3cm]{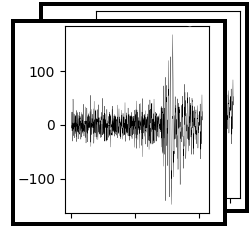}};
  \node (vae_pointcloud)[right=0cm of raw_antenna_traces]{\includegraphics[width=10cm]{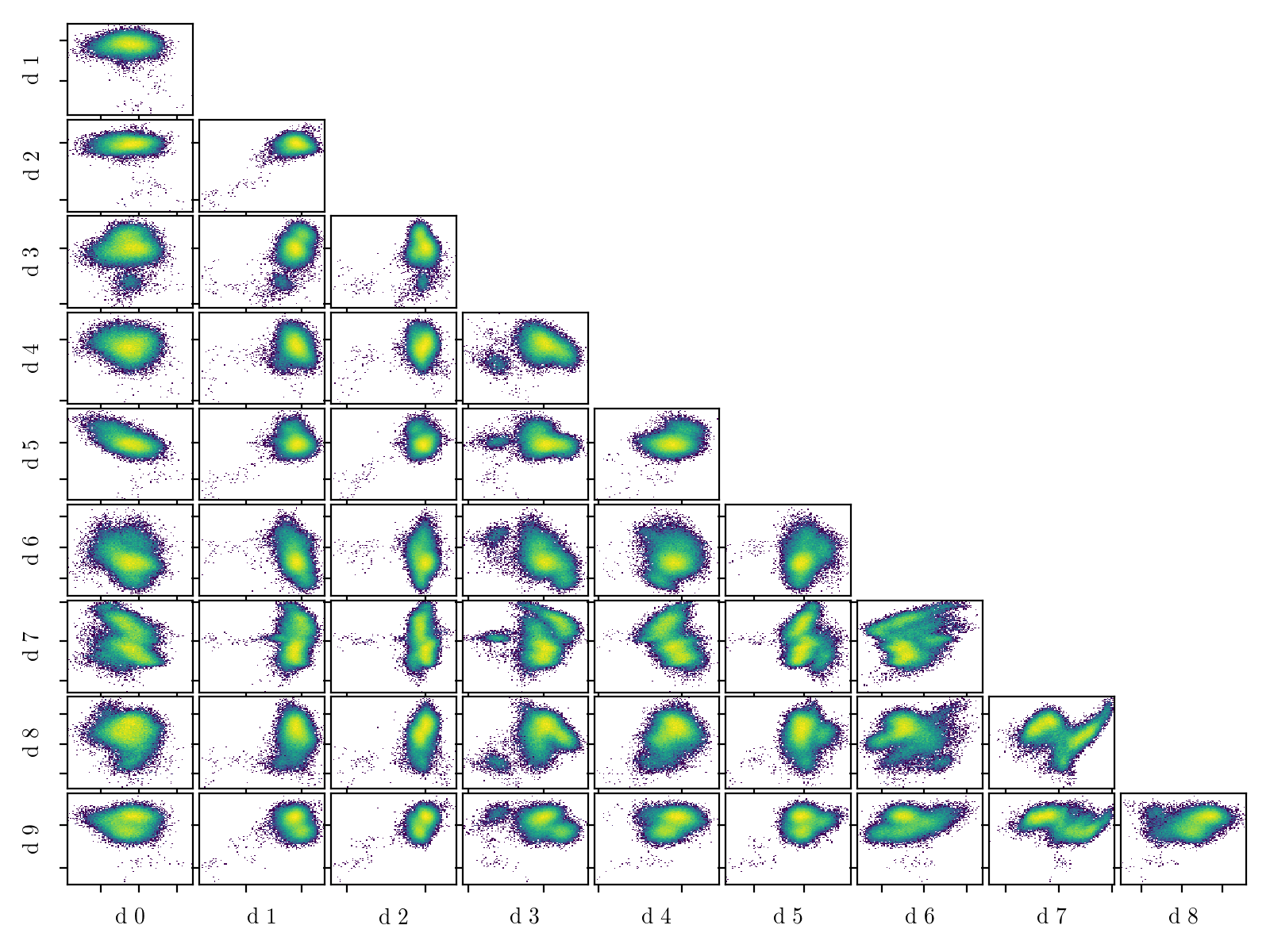}};

  \node [ above right=-6cm and -9cm of vae_pointcloud] (vae_pointcloud_with_labels) {\includegraphics[width=8.5cm]{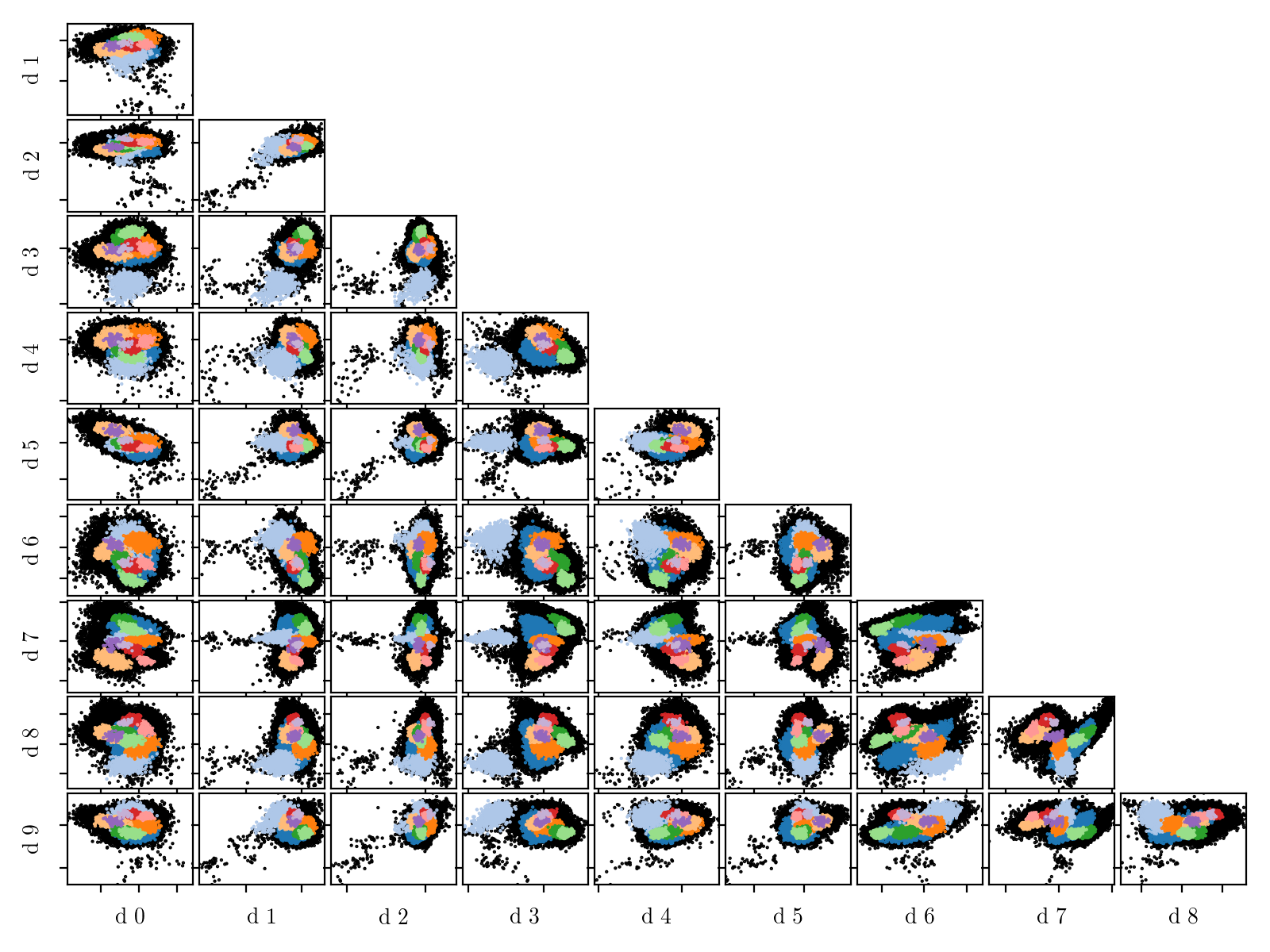}};

    \node [ below=3.75cm of raw_antenna_traces] (umap_raw_desc) {UMAP of raw data};
    \node [ below=0cm of umap_raw_desc] (umap-raw) {\includegraphics[width=7cm]{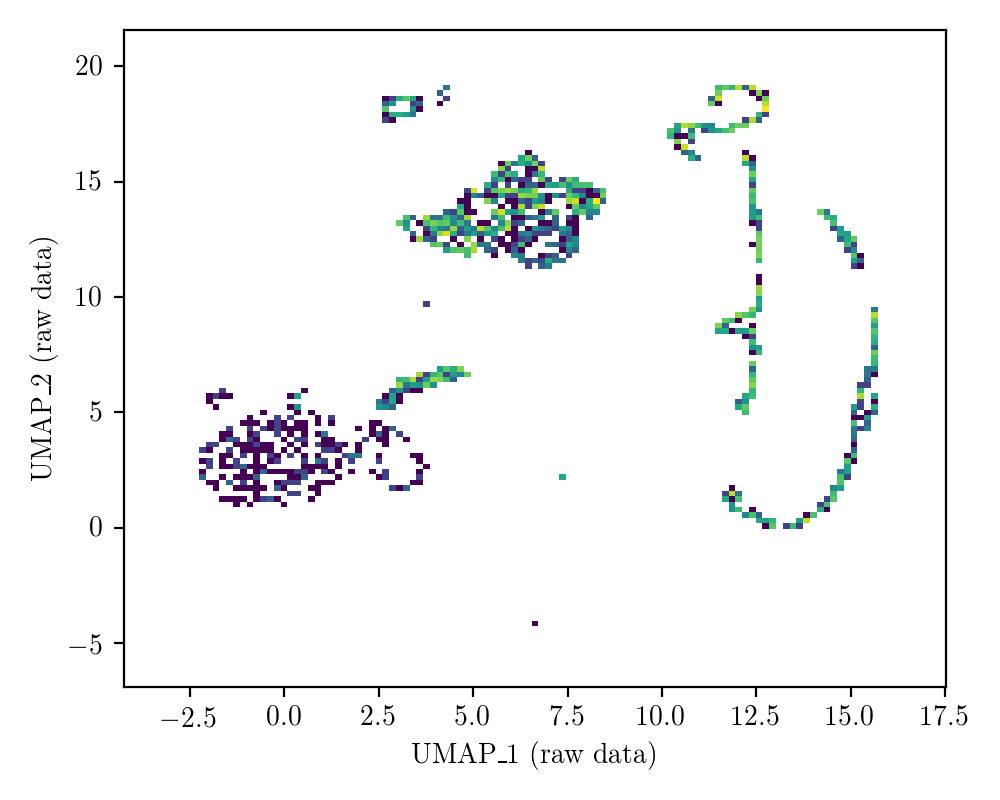}};

    \node [ below=1.5cm of vae_pointcloud] (umap_vae_desc) {UMAP of latent data};
    \node [ below=0cm of umap_vae_desc] (umap-raw) {\includegraphics[width=7cm]{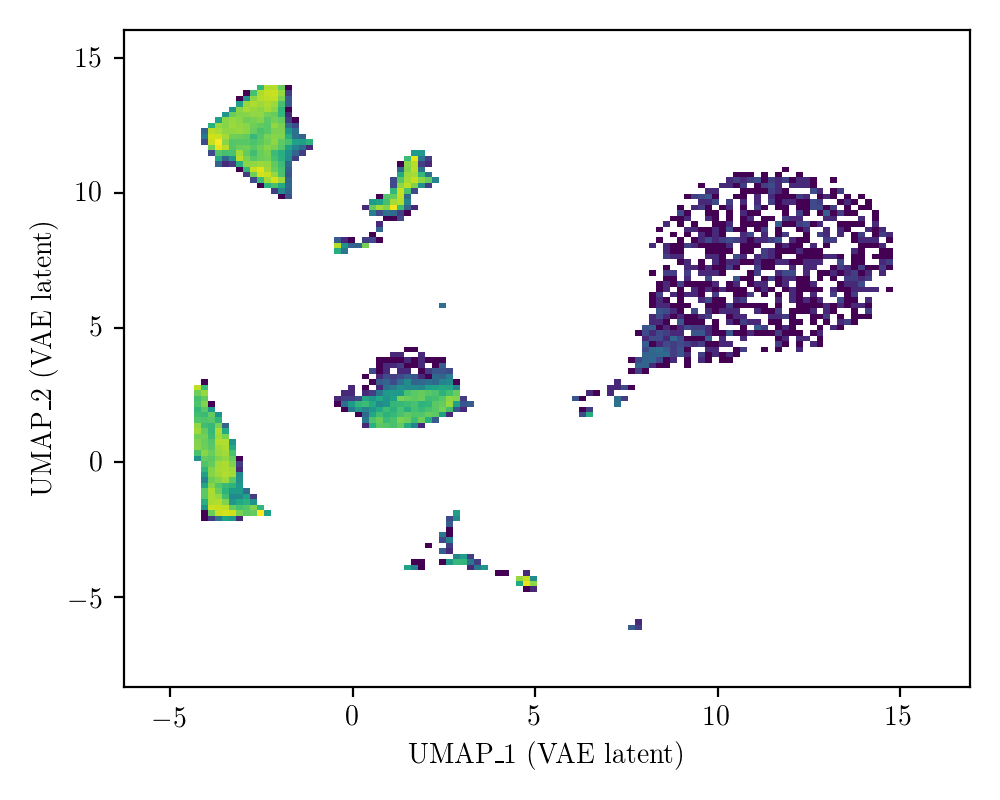}};

     \draw [->] (raw_antenna_traces.south) -- (umap_raw_desc.north);

   \draw [->] (vae_pointcloud.south) -- (umap_vae_desc.north);
\end{tikzpicture}
\caption{Overview of the clustering step. Every datum that is triggered with a voltage threshold trigger is projected into the latent space by the mean of its posterior distribution (upper right). These projections form a point cloud in the latent space onto which the clustering algorithm \texttt{H-DBSCAN} is applied (colored points). A UMAP projection of the raw data and of the latent representation are shown in the bottom. }
\label{fig:step2}
\end{figure}

\begin{figure*}[hbt]
\centering
\includegraphics[width=\textwidth]{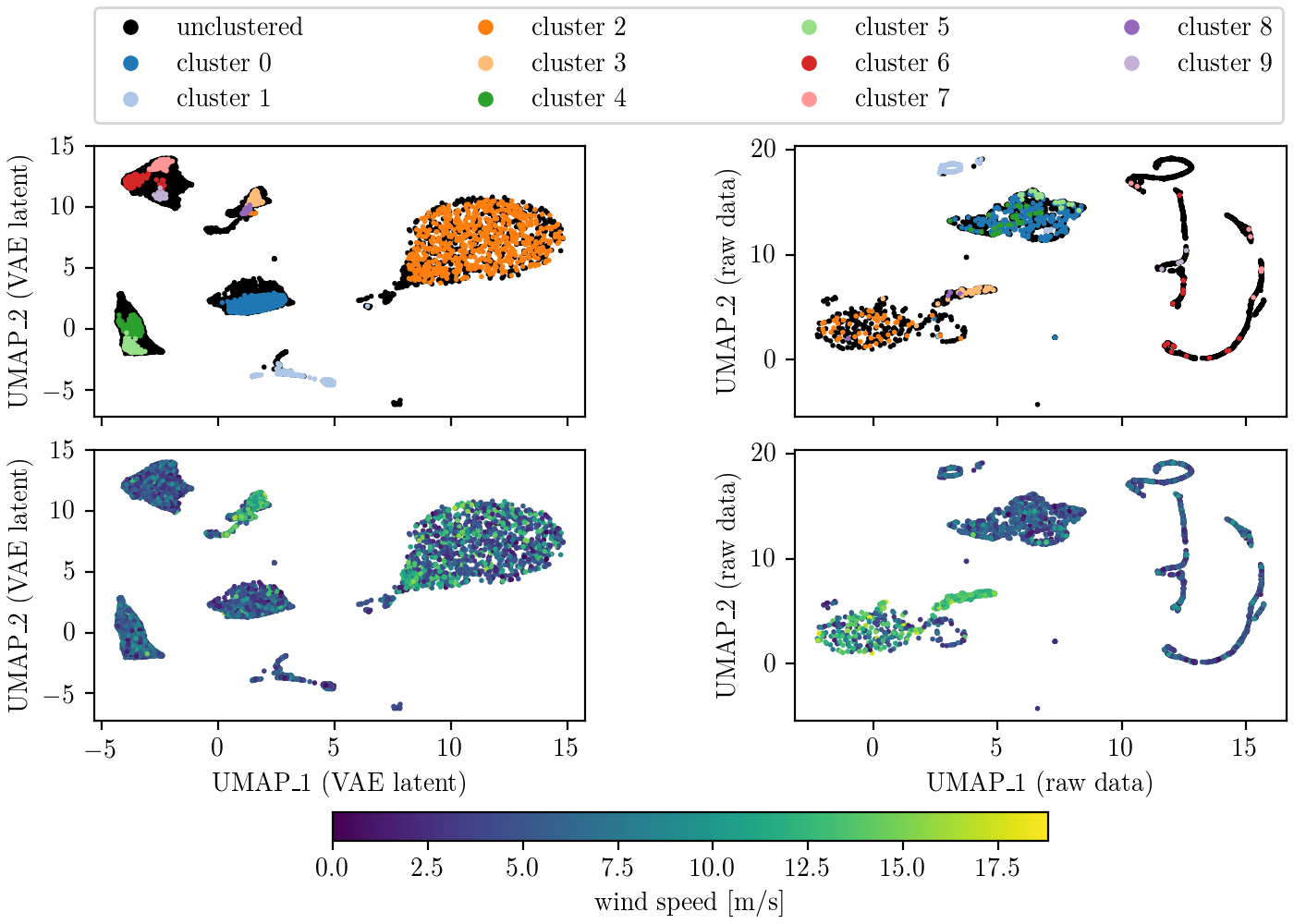}

\caption{UMAP embeddings of (noisy) station 21 events. The left column shows a VAE latent space embedding, the right column an embedding of the raw antenna data of a smaller sub-sample (20000 events). The upper row shows corresponding clustering labels, the lower row overlays the wind speed measured simultaneously to an event.}
\label{fig:outcome_21}
\end{figure*}

\begin{figure*}[hbt]
\centering
\includegraphics[width=\textwidth]{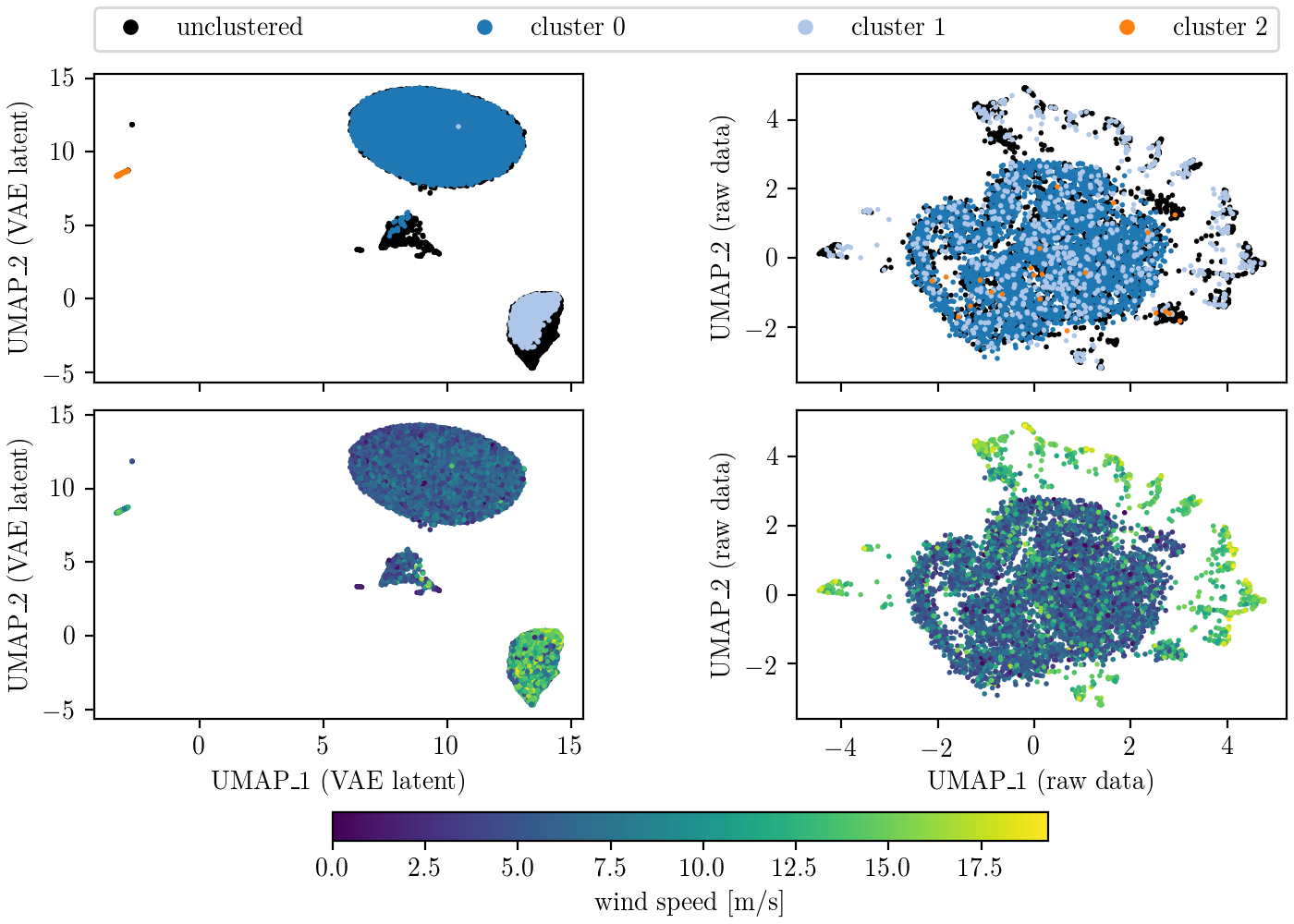}

\caption{UMAP embeddings of (silent) station 23 events. The left column shows a VAE latent space embedding, the right column an embedding of the raw antenna data of a smaller sub-sample (20000 events). The upper row shows corresponding clustering labels, the lower row overlays the wind speed measured simultaneously to an event.}
\label{fig:outcome_23}
\end{figure*}

\section{The clustering architecture}

We perform the clustering as a two step process. In the first step, we employ a variational autoencoder (VAE) \cite{vae_paper} to compress the antenna data of the shallow array (9 antennas) into a low-dimensional latent representation. A visual summary of the overall VAE processing is depicted in fig.~\ref{fig:vae_pipeline}. The VAE broadly consists of two distributions: a stochastic mapping from the antenna data to latent space, in the following called the latent posterior distribution $q_\phi(z;x)$, and a stochastic mapping from latent to antenna data, in the following called the reconstruction term $p_\theta(x;z)$. As described further below and as depicted in fig.~\ref{fig:vae_pipeline}, we actually use different representations of the antenna data for the forward and backward mapping. The parameters $\phi$ and $\theta$ are the parameters of multi-layer perceptrons\cite{dl_book} (MLPs) that map from the conditional input to the parameters of the respective distribution. In order to learn $q_\phi(z;x)$ and $p_\theta(x;z)$, we minimize the negative evidence lower bound (ELBO) as a loss function $\mathcal{L}$.
\begin{equation}
\mathcal{L}=-E_x \underbrace{\left[ \int\limits_{z}^{} q_\phi(z;x) \cdot \left[  \mathrm{ln} p_\theta(x;z) - \mathrm{ln}\frac{q_\phi(z;x)}{\mathcal{N}(z;\bm{0},\mathbbm{1})}  \right] dz \right]}_{\mathrm{evidence \ lower \   bound \ (ELBO)}}
\end{equation}
The total reconstruction term $p_\theta(x;z)$ is defined as 
\begin{equation}
    p_\theta(x;z) = \prod_{i=1}^9 p_\theta(x_i;z, \eta_i),
\end{equation} which is a product of nine probability distributions (PDFs), one for each antenna. Each individual term is modeled via a 2048 dimensional Gaussian with diagonal covariance and a shared variance across all dimensions to model a probability distribution over the observable space. Instead of the normal antenna voltage trace, we predict the logarithm of the absolute trace, which is easier to model. In addition to the latent input, we also add position and rotation information $\eta_i$ of a given antenna to describe the individual antenna responses. The target latent space $z$ is 10-dimensional for all experiments. The MLP that predicts the parameters of $q_\phi(z;x)$ takes as input the antenna data in a frequency representation of the voltage trace. We use a scattering transform \cite{scattering_transform} for this frequency representation, because it allows us to consider time-dependence. We omit details for brevity but refer to \cite{scattering_transform} for details. The output of the scattering transform are a set of coefficients, starting at 0th order, and going up to a maximal order defined by the user. We choose to go to a maximum order of two, which for our setting yields 16 0th order coefficients, 1264 1st order coefficients, and 3840 2nd order coefficients.

These are flattened to serve as a 5120 dimensional input to a MLP. This MLP is applied to all nine antennas and the output is concatenated and mapped via another MLP to $q_\phi(z;x)$. The prior distribution in the VAE is a standard normal distribution.

After training the VAE we perform clustering in the latent space. First, we generate points in the latent space by calculating the sample mean of the posterior distribution associated to each data point that has a voltage threshold trigger. The threshold trigger condition is used to filter out data that is just thermal noise, which mostly comes from other triggers. It is important to emphasize that the VAE training was performed on all triggered events, including an interval trigger that just takes data in fixed intervals. We then use \texttt{H-DBSCAN} \cite{hdbscan}, a common density based clustering algorithm, to cluster those physics-triggered events into $N$ clusters in the latent space, where $N$ is automatically determined by the algorithm. A visualization of the latent space and a colored representation including the associated cluster labels is show in fig.~\ref{fig:step2} in the upper right.
The clustering algorithm has a few tunable parameters, for which we take the standard settings. The only one we fix is the minimum cluster size, which we set to $30$, and the final cluster selection method, which we set to \texttt{leaf}.  Some data points also can not be associated to any cluster, which we call "unclustered". Finally, we apply UMAP \cite{umap} to a subsample of the raw antenna data (fig.~\ref{fig:step2} left) and to the latent representation (fig.~\ref{fig:step2} right) in order to obtain 2-d visualizations of the high-dimensional input. UMAP preserves local manifold structure and allows to qualitatively judge the clustering performance by super-imposing the cluster labels. Having two different UMAP projections allows us to better judge the performance of the VAE.

\section{Application to RNO-G data and results}

For both stations we train the setup on roughly $200000$ triggered events taken between July and October in 2022. After applying the VAE, the clustering algorithm, and the UMAP projection, we obtain the visualizations in fig.~\ref{fig:outcome_21} and fig.~\ref{fig:outcome_23} for the noisy and silent station, respectively. In the noisy station the algorithm finds 10 clusters. To understand these clusters, we use the event time distributions, the measured wind speed and a visual inspection of individual events. Clusters 0, 4 and 5 can be identified as LoRaWAN noise, sometimes overlaid with continuous wave signals. Clusters 6, 7 and 9 can be identified as battery charging noise, which appears as a 1-d structure in the raw-data UMAP projection (fig.~\ref{fig:outcome_21} top right). This comes from the fact that the underlying noise signal is always similar, just shifted due to a random triggering offset. There are two more 1-d periodic clusters which correspond to periodic trace signals from human communication (cluster 1) and the same but overlaid with LoRaWAN noise (cluster 0). Because the trigger threshold is rather high in this station, the thermal noise event rate of threshold-triggered events is rather low. We identify cluster 2 to partially contain noise events. Finally, clusters 3, 8 and also in parts cluster 2 are wind-induced events, which can be seen with the association to wind speeds (lower row in fig.~\ref{fig:outcome_21}).

\begin{table}
\begin{center}
\begin{tabular}{ c || c c  } 
 \hline
   class & \multicolumn{2}{c}{cluster ids}    \\ 
   description &  station 21 & station 23  \\
 \hline\hline
  thermal noise & 2 & 0 \\
  wind induced &  2,3,8 & 1,2 \\ 
  battery charging (+ CW)  & 6,7,9 & \\
  LoRaWAN (+ CW)  & 0,4,5 & \\
  human made (CW / pulsing)  & 1 & \\
 \hline
\end{tabular}
\caption{Overview of the different event classes found in triggered data. The cluster ids correspond to the cluster ids in fig.~\ref{fig:outcome_21} and fig.~\ref{fig:outcome_23}. CW stands for "continuous wave" signal, i.e. a single frequency signal.}
\label{tab:noise_classes}
\end{center}
\end{table}

For the silent station the algorithm detects only three clusters (fig.~\ref{fig:outcome_23}). In contrast to the noisy station there is much more thermal noise (cluster 0), because the trigger threshold is lower. Additionally, the algorithm identifies wind-events (cluster 1, 2), which can be associated due to higher wind speeds. Interestingly, in the raw-data UMAP representation, high wind speeds are situated further out in proportion to the speed, which can be seen in the lower right plot in fig.~\ref{fig:outcome_23}. This structure is not seen in the VAE-based projection, and suggests that the VAE latent space representation still has lots of potential for improvement.

An overview of the classes and associated clusters is summarized in table~\ref{tab:noise_classes}.
Ideally, the clustering settings are such that each class is associated only to a single cluster. We chose the \texttt{leaf} option in \texttt{H-DBSCAN}, which usually captures all relevant clusters, with the downside that it sometimes splits a class into multiple subclasses. A better latent representation and finetuning of the clustering algorithm should help to obtain more quantitative results in this regard, wich hopefully leads to better 1-to-1 class association in the future.

\section{Conclusion and outlook}
In this contribution, we applied a variational autoencoder to RNO-G shallow station data of a silent and noisy station. Subsequently, we performed a clustering in latent space using \texttt{H-DBSCAN}. UMAP projections of the final outcome show that the clustering makes sense, and identifies major event classes, including wind-induced background events, even on the noisy station. However, the results are qualitative, as we had to use loose clustering settings in order to capture all classes. Nontheless, the results are promising. A better latent representation and a tuning of clustering parameters are avenues for further research to obtain a better classification, and thereby a better understanding of the detector and its various background events. The method has the potential to be developed into an online monitoring tool in the future.

\bibliographystyle{ICRC}
\setlength{\bibsep}{1.pt}
\bibliography{main}
\newpage
\section*{Full Author List: RNO-G Collaboration}
\scriptsize
\noindent
J. A. Aguilar$^{1}$, 
P.~Allison$^{2}$, 
D.~Besson$^{3}$, 
A.~Bishop$^{10}$, 
O.~Botner$^{4}$, 
S.~Bouma$^{5}$, 
S.~Buitink$^{6}$, 
W.~Castiglioni$^{8}$, 
M.~Cataldo$^{5}$, 
B.~A.~Clark$^{7}$, 
A.~Coleman$^{4}$, 
K.~Couberly$^{3}$, 
P.~Dasgupta$^{1}$, 
S.~de Kockere$^{9}$, 
K.~D.~de Vries$^{9}$, 
C.~Deaconu$^{8}$, 
M.~A.~DuVernois$^{10}$, 
A.~Eimer$^{5}$, 
C.~Glaser$^{4}$, 
T.~Gl{\"u}senkamp$^{4}$, 
A.~Hallgren$^{4}$, 
S.~Hallmann$^{11}$, 
J.~C.~Hanson$^{12}$, 
B.~Hendricks$^{14}$, 
J.~Henrichs$^{11, 5}$, 
N.~Heyer$^{4}$, 
C.~Hornhuber$^{3}$, 
K.~Hughes$^{8}$, 
T.~Karg$^{11}$, 
A.~Karle$^{10}$, 
J.~L.~Kelley$^{10}$, 
M.~Korntheuer$^{1}$, 
M.~Kowalski$^{11, 15}$, 
I.~Kravchenko$^{16}$, 
R.~Krebs$^{14}$, 
R.~Lahmann$^{5}$, 
P.~Lehmann$^{5}$, 
U.~Latif$^{9}$, 
P.~Laub$^{5}$, 
C.-H. Liu$^{16}$, 
J.~Mammo$^{16}$, 
M.~J.~Marsee$^{17}$, 
Z.~S.~Meyers$^{11, 5}$, 
M.~Mikhailova$^{3}$, 
K.~Michaels$^{8}$, 
K.~Mulrey$^{13}$, 
M.~Muzio$^{14}$, 
A.~Nelles$^{11, 5}$, 
A.~Novikov$^{19}$, 
A.~Nozdrina$^{3}$, 
E.~Oberla$^{8}$, 
B.~Oeyen$^{18}$, 
I.~Plaisier$^{5, 11}$, 
N.~Punsuebsay$^{19}$, 
L.~Pyras$^{11, 5}$, 
D.~Ryckbosch$^{18}$, 
F.~Schl{\"u}ter$^{1}$, 
O.~Scholten$^{9, 20}$, 
D.~Seckel$^{19}$, 
M.~F.~H.~Seikh$^{3}$, 
D.~Smith$^{8}$, 
J.~Stoffels$^{9}$, 
D.~Southall$^{8}$, 
K.~Terveer$^{5}$, 
S. Toscano$^{1}$, 
D.~Tosi$^{10}$, 
D.~J.~Van Den Broeck$^{9, 6}$, 
N.~van Eijndhoven$^{9}$, 
A.~G.~Vieregg$^{8}$, 
J.~Z.~Vischer$^{5}$, 
C.~Welling$^{8}$, 
D.~R.~Williams$^{17}$, 
S.~Wissel$^{14}$, 
R.~Young$^{3}$, 
A.~Zink$^{5}$
\\ 
\\ 
\noindent
$^1$ Universit\'e Libre de Bruxelles, Science Faculty CP230, B-1050 Brussels, Belgium \\ 
$^2$ Dept.\ of Physics, Center for Cosmology and AstroParticle Physics, Ohio State University, Columbus, OH 43210, USA \\ 
$^3$ University of Kansas, Dept.\ of Physics and Astronomy, Lawrence, KS 66045, USA \\ 
$^4$ Uppsala University, Dept.\ of Physics and Astronomy, Uppsala, SE-752 37, Sweden \\ 
$^5$ Erlangen Center for Astroparticle Physics (ECAP), Friedrich-Alexander-Universit\"at Erlangen-N\"urnberg, 91058 Erlangen, Germany \\ 
$^6$ Vrije Universiteit Brussel, Astrophysical Institute, Pleinlaan 2, 1050 Brussels, Belgium \\ 
$^7$ Department of Physics, University of Maryland, College Park, MD 20742, USA \\ 
$^8$ Dept.\ of Physics, Enrico Fermi Inst., Kavli Inst.\ for Cosmological Physics, University of Chicago, Chicago, IL 60637, USA \\ 
$^9$ Vrije Universiteit Brussel, Dienst ELEM, B-1050 Brussels, Belgium \\ 
$^{10}$ Wisconsin IceCube Particle Astrophysics Center (WIPAC) and Dept.\ of Physics, University of Wisconsin-Madison, Madison, WI 53703,  USA \\ 
$^{11}$ Deutsches Elektronen-Synchrotron DESY, Platanenallee 6, 15738 Zeuthen, Germany \\ 
$^{12}$ Whittier College, Whittier, CA 90602, USA \\ 
$^{13}$ Dept.\ of Astrophysics/IMAPP, Radboud University, PO Box 9010, 6500 GL, The Netherlands \\ 
$^{14}$ Dept.\ of Physics, Dept.\ of Astronomy \& Astrophysics, Penn State University, University Park, PA 16802, USA \\ 
$^{15}$ Institut für Physik, Humboldt-Universit\"at zu Berlin, 12489 Berlin, Germany \\ 
$^{16}$ Dept.\ of Physics and Astronomy, Univ.\ of Nebraska-Lincoln, NE, 68588, USA \\ 
$^{17}$ Dept.\ of Physics and Astronomy, University of Alabama, Tuscaloosa, AL 35487, USA \\ 
$^{18}$ Ghent University, Dept.\ of Physics and Astronomy, B-9000 Gent, Belgium \\ 
$^{19}$ Dept.\ of Physics and Astronomy, University of Delaware, Newark, DE 19716, USA \\ 
$^{20}$ Kapteyn Institute, University of Groningen, Groningen, The Netherlands \\ 

\subsection*{Acknowledgments}

\noindent
We are thankful to the staff at Summit Station for supporting our deployment work in every way possible. We also acknowledge our colleagues from the British Antarctic Survey for embarking on the journey of building and operating the BigRAID drill for our project.
We would like to acknowledge our home institutions and funding agencies for supporting the RNO-G work; in particular the Belgian Funds for Scientific Research (FRS-FNRS and FWO) and the FWO programme for International Research Infrastructure (IRI), the National Science Foundation (NSF Award IDs 2118315, 2112352, 211232, 2111410) and the IceCube EPSCoR Initiative (Award ID 2019597), the German research foundation (DFG, Grant NE 2031/2-1), the Helmholtz Association (Initiative and Networking Fund, W2/W3 Program), the University of Chicago Research Computing Center, the European Research Council under the European Unions Horizon 2020 research and innovation programme (grant agreement No 805486) and the Carl Tryggers foundation.

\end{document}